\documentclass[12pt]{iopart}
\usepackage{amsfonts}
\usepackage{amsthm}
\usepackage{amscd}
\usepackage{ulem}
\usepackage{cancel}
\usepackage{amssymb}
\usepackage{graphicx}
\usepackage{color}
\usepackage{graphics}

\newcommand{\half}{{\textstyle \frac{1}{2}}}

\newcommand{\uvec}{{\bf u}}

\theoremstyle{definition}

\theoremstyle{remark}

\newcommand{\dst} {\displaystyle}

\newcommand{\la} {\lambda}

\newcommand{\eps} {\epsilon}

\newcommand{\lb} {\left(}
\newcommand{\rb} {\right)}
\newcommand{\lbr} {\left\{}
\newcommand{\rbr} {\right\}}

\newcommand{\rd} {\right.}

\newcommand{\bga}{\begin{array}{l}}
\newcommand{\ena}{\end{array}}
\newcommand{\bge}{\begin{equation}}
\newcommand{\bgea}{\begin{equation} \begin{array}{l} }
\newcommand{\ene}{\end{equation}}
\newcommand{\enea}{ \end{array} \end{equation}}

\begin{document}
\title{Simulation of the Elastic Properties of Reinforced Kevlar-Graphene Composites}

\author{ Karen S. Martirosyan$^\dag$,  M. Zyskin$^{\sharp}$,}
%\address{$^1$ Department of Mathematics}
\address{$^\dag$ Department of Physics and Astronomy,  
University of Texas at Brownsville,  \\
80 Fort Brown, Brownsville,
TX 78520, USA}
\address{$^\sharp$Rutgers University, 126 Frelinghuysen Road, Piscataway, NJ 08854-8019, USA} 
\ead{Karen.Martirosyan@utb.edu, zyskin@yahoo.com}

\begin{abstract}

The compressive strength of unidirectional fiber composites in the form of Kevlar yarn with a thin outer layer of graphene was investigated and modeled. Such fiber structure may be fabricated by using a strong chemical bond between Kevlar yarn and graphene sheets. Chemical functionalization of graphene and Kevlar may achieved by modification of appropriate surface-bound functional (e.g., carboxylic acid) groups on their surfaces. In this report we studied elastic response to unidirectional in-plane applied load with load peaks along the diameter. The 2D linear elasticity model predicts that significant strengthening occurs when graphene outer layer radius is about 4 \% of kevlar yarn radius. The polymer chains of Kevlar are linked into locally planar structure by hydrogen bonds across the chains, with transversal strength considerably weaker than longitudinal one. This suggests that introducing outer enveloping layer of graphene, linked to polymer chains by strong chemical bonds may significantly strengthen Kevlar fiber with respect to transversal deformations.

\end{abstract}
\pacs{62.20.-x, 68.35.Gy, 81.40.-z }

%\ams{74B10, 74G05, 74G55}
\maketitle

\section{Introduction}

Reinforced polymer fibers have been studied intensively in search of superior fabrics for a broad range of applications including bullet-proof vests, protective clothing, and high-performance composites for aircraft and automotive industries  \cite{1,2} %[1-2]
. For most structural materials, the compressive strength is much greater than the tensile strength. Fiber-reinforced composites are among a few materials that exhibit greater tensile strength than compressive strength. This behavior results from fiber micro buckling compressive failure mechanism in fiber composites. Static compressive strengths of unidirectional composites have been studied by several research groups in the past few decades
[3-6].

A major step forward in the protection of personnel against ballistic threats has been achieved by introduction of Kevlar synthetic fibers and textiles. The most common approaches for the fabrication of reinforced composite fibers have been melt processing and solution coagulation spinning \cite{7,8}
%[7, 8].
Kevlar fibers are produced by wet-spinning from sulfuric acid solutions \cite{9}.
%[9].
While in acid solution, Kevlar form liquid crystalline phase, with rod-like polymer chains  well-aligned parallel to each other, and bonded by hydrogen bonds across to form strong planar sheets [10]. Such sheets are stacked radially to form Kevlar fiber. As a result, the fiber is much stronger with respect to longitudinal deformations than to transversal deformations.

On the other hand, the carbon nanotubes (CN) and graphene sheets (GS) have been proposed as one of the most promising additives for the fabrication of ultra-strong polymer composites due to their advanced mechanical properties \cite{11,12}.
%[11, 12].
It is well-known that carbon nanotubes and graphene sheets have Young's modulus and tensile strength above 1 TPa and 60 GPa,
[13-16]
%\cite{13}-\cite{15_2}
respectively, while their densities can be as low as 1.3 $\mbox{g/cm}^3$.  They may exist as single layered or multi-layer structures. It is possible to harness the multifunctional properties of graphene sheets and design novel class of advanced composites with superior mechanical and electric performance
\cite{16,17}
%[16, 17].
Thus, various polymeric-CN and polymeric-GS composites and their chemical functionalisations have become a subject of intensive research and technological development over the last years
\cite{18,19}.
%[18, 19].
Recently, a significant mechanical enhancement of Kevlar fibers has been demonstrated  by incorporation of surface modified carbon nanotubes achieved by using N-methylpyrrolidone (NMP) solvent
\cite{20,21}.
%[20, 21].
The maximum values observed were for Young's modulus 115-207 GPa; yield strength 4.7-5.9 GPa; strain at break 4.0-5.4 \%; and toughness 63-99 J/g
\cite{20}.
%[20].

In this paper we present the 2D linear elasticity model to predict the mechanical properties of the  Kevlar fibers reinforced by depositing an outer shell of graphene, rigidly attached to Kevlar fiber by chemical bonding. The schematic diagram of the kevlar-graphene fibers is shown in Figure 1,(a). Graphene outer layer can be formed by soaking kevlar single fiber with graphene dispersed in N-methylpyrrolidone (NMP) solvent. By using spinning technique it will be possible to create a composite structure by depositing an outer layer of graphene enveloping the Kevlar fiber. This lead us to hypothesize that graphene could be incorporated into swelled Kevlar fibers\ by soaking fibers in a dispersion of graphene in NMP. Chemical functionalization of graphene and Kevlar may be achieved by modification of appropriate surface-bound functional (e.g., carboxylic acid) groups on their surfaces. We anticipate that outer graphene layer enveloping Kevlar fiber may stabilize radially stacked hydrogen-bonded planar sheets forming  Kevlar fiber with respect to transversal deformations.

The main goal of this paper is to model the mechanical behavior of kevlar-graphene fibers and to predict the optimal thickness of the outer layer of graphene for strengthening. We show that within 2d linear elasticity model, and taking realistic Young modulus and Poisson ratios for transversal deformations, reinforcement starts to occur when the outer hard shell radius is about 4 \% of kevlar fiber radius (corresponding to about 8 \% weight). We assume that  applied load has  a narrow load peak along the diameter, which we may take to be the Y axes; typical displacements are shown in Figure 1,(b).

Our modeling assumes that graphene flakes are tied up together by covalent bonds, which may be achieved by graphene functionalization. In the presence of strong interlayer covalent bonds, we assume that bonded graphene flakes will have high Young modulus and tensile strength in all directions, and we model it by isotropic continuous model.  Such model predicts theoretical limit on strengthening which may be achieved.

Commercially available Kevlar fiber has radius of about 6 microns.  In that case, we will need about 240 nanometers of functionalized multilayer graphene for substantial two-fold yield strength improvement. At such scale, continuous elasticity model may be used.

\section{Explicit solution}
We assume that the enveloping graphene layer and Kevlar fiber are rigidly attached by chemical bonds on their common boundary, and that the  load is perpendicular to the fibers.

In this section we derive an explicit solution of planar linear elasticity equations, modeling  a circular fiber enveloped by a rigidly attached to it ring of different stronger material.

In section 3, we use our explicit solution to compute stress tensor, both in the outer graphene layer and inside Kevlar fiber, and to analyze increase of the yield strength of the composite system, as a function of reinforcing layer thickness.

Our explicit solution is given in terms of  the Airy stress function $U.$  Airy stress function is a function from which the stress (and strain) tensors  can be derived at any given point
$(x, y);$  those  stress and strain tensors will then automatically satisfy the equilibrium equations of linear elasticity. Airy stress function is a solution of bi-harmonic equation
\bge
\Delta^2 U =0 .
\ene
It is related to the stress tensor  by

\bge
\bga
\sigma_{yy} =\frac{\partial ^2 U }{\partial  x^2 } ,
\\
\sigma_{xx} =\frac{\partial ^2 U }{\partial  {y^2 }},
\\
\sigma_{xy} =-\frac{\partial ^2 U }{\partial {x } \partial {y }} .
\label{StressTensorFromU}
\ena
\ene

It is well-known \cite{22} that a biharmonic function can be expressed via two analytic functions $\varphi (z) $,  $\chi (z)$,  $z= x+ i y$  :
\bge
2 U = \bar{z} \varphi+ z \bar{\varphi} + \chi + \bar{\chi} .
\ene
It follows from (\ref{StressTensorFromU}) that
\bge\bga
\sigma_{xx} + \sigma_{yy} =
% 2 \lb \phi^\prime +  \bar{\phi}^\prime \rb ,
2 \lb \Phi  +  \bar{\Phi}  \rb ,
\\
\sigma_{yy}-\sigma_{xx}  + 2 i \sigma_{xy} = 2 \lb \bar{z} \Phi^\prime + \Psi \rb, \quad \Phi \equiv \varphi^\prime, \Psi \equiv \chi^{\prime \prime} .
\ena\ene
Changing variables to polar coordinates $(r, \theta)$, we have
\bge\bga
\sigma_{rr} + \sigma_{\theta \theta} =  2 \lb \Phi  +  \bar{\Phi}  \rb ,
\\
\sigma_{\theta \theta}-\sigma_{rr}  + 2 i \sigma_{r\theta} = 2 e^{2 i \theta} \lb \bar{z} \Phi^\prime + \Psi \rb,
\label{StressPolar}
\ena\ene
and it follows that
\bge\bga
\sigma_{rr} -i \sigma_{r\theta }  = \Phi  +  \bar{\Phi}  - e^{2 i \theta} \lb \bar{z} \Phi^\prime + \Psi \rb .
\label{sig_rr-i sig_rth}
\ena\ene
For the displacement $\uvec= (u_x, u_y) $, it follows from the Hook's Law that
\bgea
2 \mu \lb u_x + i u_y \rb =
2 \mu \lb u_r + i u_\theta\rb e^{i \theta} =
 - 2 \bar{\partial} U + 2 \dst\frac{\la + 2 \mu }{\la + \mu } \varphi =
\kappa \varphi - z \bar{\varphi}^\prime - \bar{\chi}^\prime ,
\\
\kappa   = \dst\frac{ \lambda  +3 \mu }{ \lambda + \mu  }
\label{Displ}
\enea
(equation (\ref{Displ}) goes back to Lam\'{e}; the derivation can be found in Ref (22), \S 32).
Lam\'{e} parameters $\la$, $\mu$ are related to the Young modulus $E$ and Poisson ratio $\nu$ as follows,
\bgea
\la =   \frac{E \nu}{(1+\nu) ( 1- 2 \nu)},\mu = \frac{E}{2 (1+\nu)} .
\label{LameYoung}
\enea

We expand functions $\Phi$, $\Psi$ in the Laurent  series in the outer ring, and in Taylor series on the inner disk,
\bgea
\underline{R_1 \leq \vert z \vert \leq R_2}
\\
\Phi = \dst \mathop{\sum}_{n=-\infty}^{\infty} a_n \dst \lb \frac{z}{R_2} \rb^n ,
%\\
\Psi = \dst \mathop{\sum}_{n=-\infty}^{\infty} p_n \dst \lb \frac{z}{R_2} \rb^n ;
\label{SolutionOuter}
\enea
\bgea
\underline{\vert z \vert \leq R_1 }
\\
\Phi = \dst \mathop{\sum}_{n= 0}^{\infty} b_n \dst \lb \frac{z}{R_2} \rb^n ,
%\\
\Psi = \dst \mathop{\sum}_{n= 0}^{\infty} q_n \dst \lb \frac{z}{R_2} \rb^n .
\label{SolutionInner}
\enea
%For convenience, we take $b_n = q_n = 0 , n<0$.
We assume that  the external stress applied to  the outer boundary    $\vert z \vert = R_2$ is known, and has a Fourier series expansion
\bge\bga
\sigma_{rr} -i \sigma_{r\theta } =  \dst \mathop{\sum}_{n=-\infty}^{\infty} A_n
e^{i n   \theta} .
\label{outerStress}
\enea
Thus on the external boundary $\vert z \vert = R_2$ from (\ref{sig_rr-i sig_rth}), (\ref{SolutionOuter}), (\ref{outerStress})   we have that
\bgea
(1-n) a_n
 -p_{n-2}+ \bar{a}_{-n}
  =
  A_{n} .
\label{External}
\enea

We now turn to analyzing conditions on the internal boundary  $\vert z \vert = R_1.$
Let us introduce dimensionless radius $r_1,$
\bgea
r_1 \equiv \dst\frac{R_1}{R_2},
\enea
and let us use the convention that
\bgea
b_n = q_n = 0 , n<0 .
\label{bn qn n<0}
\enea
On the internal boundary $\vert z \vert = R_1, $ stress components $\sigma_{rr}, $ $\sigma_{r\theta }$ and the displacement are continuous. Using  (\ref{sig_rr-i sig_rth})-(\ref{SolutionInner}), continuity of such stress components imply that
\bgea
(1-n) a_n r_1^2  -p_{n-2}+ \bar{a}_{-n} r_1^{2-2 n}  = (1-n) b_n r_1^2  -q_{n-2}+ \bar{b}_{-n} r_1^{2-2 n},
\label{InternalStress}
\enea
and the  continuity of displacement imply that
\bgea
\dst\frac{1}{\mu_2} \lbr (n-1)  r_1^2 a_n + p_{n-2}+ \kappa _2 \bar{a}_{-n}  r_1^{-2 n +2} \rbr =
 \dst\frac{1}{\mu_1} \lbr (n-1)  r_1^2 b_n + q_{n-2}+ \kappa _1 \bar{b}_{-n}  r_1^{-2 n +2} \rbr .
 \label{InternalDispl}
\enea
Using (\ref{External}) to eliminate $p_n,$
\bgea
\hspace{-10mm} -(n-1) \left(1-r_1^2\right) a_{n}-(n-1) r_1^2 b_{n}-{q}_{n-2}+\left(1-r_1^{2-2 n}\right) \bar{a}_{-n}+r_1^{2-2 n} \bar{b}_{-n} =   A_{n} ,
\label{neq1}
\enea
\bgea
\hspace{-10mm} -(n-1) \left(1-r_1^2\right) a_{n} +\bar{a}_{-n} \left(1 + r_1^{2-2 n} \kappa_2\right)
-\frac{\mu _2}{\mu _1}\lbr (n-1) r_1^2 b_{n} +  {q}_{n-2}  + r_1^{2-2 n} \kappa_1 \bar{b}_{-n}  \rbr \\
 =  A_{n}
\label{neq2}
\enea
We can solve (\ref{neq1}), (\ref{neq2}), supplemented with complex conjugates of those equations. We denote complex conjugates of coefficients by $\bar{a}, \bar{b}$, etc.  Complex conjugation and    $n \rightarrow -n$ in (\ref{neq1}), (\ref{neq2}) yields

\bgea
\hspace{-10mm} \left(1-r_1^{2+2 n}\right) a_{n}+r_1^{2+2 n} b_{n} - ( 1+n) \left(1-r_1^2 \right) \bar{a}_{-n}+(1+n) r_1^2 \bar{b}_{-n}-\bar{q}_{-2-n} = \bar{A}_{-n}
\label{cneq1}
\enea
\bgea
\hspace{-10mm} \left(1+r_1^{2+2 n} \kappa _2\right)a_{n}   +(1+n) \left(1-r_1^2 \right) \bar{a}_{-n} -
\frac{\mu _2}{\mu _1}
\lbr -(1+n) r_1^2 \bar{b}_{-n} +  \bar{q}_{-2-n} +  r_1^{2+2 n}  \kappa_1 b_{n} \rbr  =\\
  \bar{A}_{-n} .
\label{cneq2}
\enea

From the zero torque condition, we have that
\bgea
 \bar{A}_{0}=A_{0} ;
 \label{A0}
\enea
this is the condition for (\ref{neq1})-(\ref{cneq2})  with $n=0$ to have solution. In fact,  solution for  $n=0$ mode is not unique.  However,  it is clear from (\ref{StressPolar}) that  imaginary part of $a_0$ does not affect the  stresses, thus we can take
\bgea
\bar{a}_{0}=a_{0}.
\enea

Assuming zero total  force,
\bgea
A_{1}=0
\label{A1};
\enea
with this condition and using (\ref{bn qn n<0}), equations (\ref{neq1}) , (\ref{neq2}) are consistent for  $n=1$ mode, and similarly equations  (\ref{cneq1}) , (\ref{cneq2}) are consistent for  $n= -1 .$

Solving equations (\ref{neq1})- (\ref{cneq2}), we can find  coefficients $ \lbr a_i, b_i, q_i \rbr , i\in \Bbb{Z} :$
%\tiny
%\small
\bgea
 a_{0} =  \half \  \frac{ 2 \mu _1- \mu _2+  \kappa _1 \mu _2   }{2 \mu _1 - \mu _2 + \kappa _1 \mu _2 + r_1^2 \lb \kappa _2 \mu_1 - \kappa _1 \mu _2 +  \mu _2  - \mu _1 \rb} A_{0},
 \\
 b_0 =   \half \  \frac{  \mu _1 +  \kappa _2 \mu _1  }{2 \mu _1 - \mu _2 + \kappa _1 \mu _2 + r_1^2 \lb \kappa _2 \mu_1 - \kappa _1 \mu _2 +  \mu _2  - \mu _1 \rb}  A_{0} ,
 \\[2mm]
 b_1 =  \frac{ \mu _1+ \kappa _2 \mu _1  }{ \mu _1+  \kappa _1 \mu _2 + r_1^4 \lb \kappa _2 \mu _1-  \kappa _1 \mu _2\rb }\bar{A}_{-1} ,
 \\
 \bar{a}_{-1} =  0,

\\

a_{1} =  \frac{  \mu _1+\kappa _1 \mu _2 }{\mu _1+  \kappa _1 \mu _2 + r_1^4 \lb \kappa _2 \mu _1-  \kappa _1 \mu _2\rb }\bar{A}_{-1},

\\[2mm]
\\

\hspace{-10mm} \left(
\begin{array}{l}
a_{n} \\
 \bar{a}_{-n} \\
 b_n \\
{q}_{n-2}
\end{array}
\right)
=
\left(
\begin{array}{cccc}
 -( n-1) \left(1-r_1^2\right) &  2 r_1^{n-1}+1 & -(n-1) r_1^2 & -1 \\
 1-r_1^{2+2 n}& (1+n) \left(1-r_1^2\right) & r_1^{2+2 n} & 0 \\
 -( n-1) \left(1-r_1^2\right) & 1+r_1^{2-2 n} \kappa _2 & \frac{-(n-1) r_1^2 \mu _2}{\mu _1} & -\frac{\mu _2}{\mu _1} \\
 1+r_1^{2+2 n} \kappa _2 & (1+n) \left(1-r_1^2\right) & \frac{- r_1^{2+2 n} \kappa _1 \mu _2}{\mu _1} & 0
\end{array}
\right)^{-1}

 \left(
\begin{array}{l}
 A_{n}\\
 \bar{A}_{-n} \\
 A_{n} \\
 \bar{A}_{-n}
\end{array}
\right),
\\
n\geq 2  , r_1 \equiv  \frac{R_1}{R_2} .
\label{coeff}
\enea

\normalsize
We note that for the displacement (\ref{Displ}) to be single-valued, we must have $\kappa _1 a_{-1} - \bar{a}_{-1} =0$, which is satisfied by  our solution (\ref{coeff}) since $a_{-1} = \bar{a}_{-1} =0.$

Coefficients $\lbr p_n \rbr  $ can be found from (\ref{External}),  (\ref{coeff})
\bgea
p_n = -  A_{2+n}-(1+n)   a_{2+n}+ \bar{a}_{-2-n} , \quad  n\in \Bbb{Z} .
\label{coeff_p}
\enea

We note that it is evident from the above formulas (as well as from dimensional analysis) that for a prescribed  applied external stress, given as a function of the polar angle, the solution for stresses on the domain depends on the ratio of internal and external radii $r_1 = \frac{R_1}{R_2},$ but not the overall scale $R_2.$

\section{Analyzing solution. }

For a given load, we can compute components of stress tensor using equations (\ref{StressPolar}), (\ref{SolutionOuter}), (\ref{SolutionInner}), (\ref{bn qn n<0}),
 (\ref{coeff}), (\ref{coeff_p}); those formulas determine stresses created by any transversal load.

In this paper we will analyze in detail response to uni-directional transversal compression applied along a diameter.
Such load may be used to estimate a response to the transversal compression of a fiber created by impacting   projectile,  when the fiber is pressed against a matrix and other fibers to stay in place. We  assume that in such case high loads will be applied in a small neighborhood  of points of contact, and will be uni-directional. Young modulus of the reinforcing outer graphene layer is very high, thus we anticipate that material delivering external load (e.g. a projectile and the  matrix,  or loading plate in a static loading experiment)  will be deformed at a point of contact with the reinforced fiber. To estimate stresses in the Kevlar-graphene system we assume that the load is uniform in a small region of contact.

The load we use in our study is shown in Figure 2. Compressing force is applied on the  outer boundary in the $(- Y)$ direction in a small neighborhood of $Y$ axes intercept, with an opposite balancing force applied in diametrically opposite region. Magnitude of the force per transversal to Y surface area element, as a function of the polar angle $\theta,$ has narrow peaks centered at  $\theta =  \pm \dst \frac{\pi}{2};$ we take the magnitude to be constant (1 GPa ) at such peaks, and we take width of the peak to be $ \mbox{w}  = \frac{\pi}{16} .$ Outside of the peaks, we take a smooth interpolation to a zero value in a small angular region of width $\eps .$ Thus  $f_Y$, the Y axes projection of the applied force per transversal to Y area element,  as a function of polar angle  is given by
\bge
f_y \lb \theta \rb = \lbr \begin{array}{rl} -1, &  \vert \theta - \frac{\pi}{2}  \vert \leq \dst\frac{\theta_{\mbox{w}}}{2}
\\
1, & \vert \theta + \frac{\pi}{2}  \vert \leq  \dst\frac{\theta_{\mbox{w}}}{2}
\\
0, &  \vert \theta \pm  \frac{\pi}{2} \vert > \theta_{\mbox{w}} + \eps .
\ena \rd
\label{f_y}
\ene
Corresponding  stresses applied on the outer boundary , as functions of polar angle, are
\bgea
 \sigma_{rr} \lb \theta \rb  = f_y ( \theta) \vert \sin\theta\vert  \sin\theta ,
 \\
 \sigma_{r\theta } \lb \theta \rb   = f_y ( \theta)\vert \sin\theta\vert  \cos\theta ;
 \label{appliedStress}
\enea
$f_y$ as a function of the polar angle $\theta$ is shown in Figure 2 (a). In Figure 2(b) and 2 (c) we plot components of the  stress tensor $\sigma_{rr} $  and $\sigma_{ r \theta  } ,$ induced in the material by  such applied load, as a function of the polar angle; here  r,  $\theta$ are polar coordinates. Stress $\sigma_{ rr  } $ corresponds to a compression in the vicinity of two diametrically opposite points on the Y axes.  Stress $\sigma_{ r \theta  } $ rapidly  changes   sign at those points (this is because  projection of $f_y$ onto the polar unit vector ${\mathbf e}_{\theta}$ changes sign).

To understand how stresses are distributed in the Kevlar fiber and the outer graphene layer, we compute components of the stress tensor using equations (\ref{StressPolar}), (\ref{SolutionOuter}), (\ref{SolutionInner}), (\ref{bn qn n<0}), (\ref{coeff}), (\ref{coeff_p}).

We take experimental values for the transverse Young modulus $E_1 = 5$ GPa and Poisson ratio $\nu_1 = 0.35$ for  Kevlar, and $E_2 =  600$ GPa and Poisson ratio $\nu_2 =0.1$ for the graphene in our analysis, and we use the relationship (\ref{LameYoung}) to estimate the Lam\'{e} parameters $\la$ and $\mu$ of Kevlar and graphene. For the particular combination of Lam\'{e} parameters which appear in the exact solution (\ref{coeff}) we get: $\mu_1 = 1.85 \ GPa$, $\kappa_1 =  1.6 $ for the Kevlar, $\mu_2 = 270 \ GPa$, $\kappa_2 =  2.6 $ for the graphene. We note that using the five elastic constants describing anisotropic, transversally isotropic material (when those constants are known) will give more accurate determination of the transversal Lam\'{e} parameters. Those five  elastic constants were measured for the Kevlar KM2 fiber \cite{24}; using those, we get values of the same order as above for $\mu_1  $ and $\kappa_1,$ and the qualitative behavior of the solution is similar.

In Figure 3 we plot the $\sigma_{yy}$ component of the stress tensor on the Y axes (that is on the line where the load is applied), for varying values of the  ratio of the inner to outer radius $r_1= \frac{R_1}{R_2}:$ We clearly see that the stress in the Kevlar fiber region is considerably reduced, dropping down in the stronger graphene region, with about 2 fold drop when $r_1  \simeq  0.96,$ and even bigger drop for smaller  $r_1.$   The magnitude of the drop varies continuously when we vary $r_1 .$  When $r_1 $  approaches $1$ , there is no reduction of stress, as expected.

In Figure 4, we plot all 3 components of the stress tensor, in the Kevlar fiber region, for varying thickness of the outer graphene shell. When graphene shell is thin, e.g.  $r_1= \dst\frac{R_1}{R_2}  \simeq  0.99,$ there is not much reduction in stresses in Kevlar; but  for a thicker outer shell, $r_1 \simeq  0.96,$ there is about 2 fold reduction of  stresses in the Kevlar fiber region.
We see that stresses in the Kevlar fiber region attain a maximum  value in the outmost region of the fiber and near the Y axes where the load is applied. Stresses in Kevlar fiber region vary continuously with the change of $r_1$, getting smaller when the outer graphene shell gets thicker in a continuous manner, as expected.

In Figure 5, we plot the  $\sigma_{rr} $ and $\sigma_{r\theta}$ components of the stress tensor in the Kevlar fiber region, corresponing to polar coordinates $r, \theta.$ Those two components of stress tensor are selected since those are continuous across the inner boundary between Kevlar and graphene regions. Behavior of those components in the Kevlar fiber region is similar to the stress components in the Euclidean  coordinates $(x,y)$ shown in Figure 4:  stresses are larger in the outer parts of the fiber and near the $Y$ axes where the external load is applied; for a fixed external load stresses decrease continuously with decrease of $r_1= \dst\frac{R_1}{R_2}, $ with approximately 2-fold decrease at $r_1 =0.96 .$

\subsection{Estimating yield }
Yield criteria for a broad class of materials may be given in terms of the Von Mises  stress.
The Von Mises  stress is proportional to the square root of the sum of squares of  characteristic values of the trace-free part of the stress tensor. Since Kevlar fiber is transversally isotropic, and the transversal yield strength of the Kevlar fiber is much lower than the longitudinal one, we will use a 2 dimensional version of the Von Mises stress,
\bgea
\sigma_{M  } =
\sqrt{\lb \sigma_{xx}- \sigma_{yy}\rb^2+4 \sigma_{xy}^2  }.
\label{VonMisesStress}
\enea
We note that for a planar elasticity problem we consider, the  longitudinal stress $\sigma_{zz} = \frac{c_{13}}{2 (\la + \mu) } \lb \sigma_{xx}+  \sigma_{yy} \rb \sim 0.6 \lb \sigma_{xx}+  \sigma_{yy} \rb$, and so is of the same order as planar stresses, while longitudinal yield is much higher than the transversal yield. Thus using more  accurate for anisotropic material yield criteria such as  Hill yield criteria  will produce similar conclusions. %(It appears that yield constants in Hill yield criteria are not known for Kevlar.)

To illustrate how the   Von Mises stress is distributed on the Kevlar-graphene domain, we plotted the Von Mises stress (\ref{VonMisesStress}) on the whole domain, Figure 6 (a), and on the  Kevlar fiber region, Figure 6 (b), in the
case when $r_1 = \frac{R_1}{R_2} = 0.95 . $ The external applied load is along the $Y$ axes and has a sharp peak of polar angle width $\frac{\pi}{16}, $   as in (\ref{f_y}), (\ref{appliedStress} ). Distribution of the von Mises stress is largely similar  to the distribution of components of the stress tensor in Figure 4. There is a sharp drop of the Von Mises stress occuring in the stronger outer graphene layer, with a more than 2 fold decrease of stress in the Kevlar fiber compared to the case of no graphene reinforcement. On the Kevlar fiber domain, the Von Mises stress is larger in the outmost region and near the $Y$ axes where the external stress is applied.

Von Mises yield criteria is that yield occurs when $\sigma_{\mbox{\small max }},$  the maximum  value of the Von Mises stress $\sigma_{M}$ taken over the domain, becomes greater or equal to the critical value $\sigma_{\mbox{\small yield }}$, specific for the material.  For Kevlar,  $\sigma_{\mbox{\small yield }}   \approx 2.9  $ GPa for longitudinal loads \cite{10};
%[10];
but for transversal loads it is lower (e.g. for Kevlar/PEKK composite the transversal strength is 21 MPa,
\cite{23}).
%[23]).

If, at a fixed external load, the maximum  von Mises stress in the Kevlar region is reduced by a factor of 2, say, due to reinforced outer graphene layer,  that implies  that combined Kevlar-graphene system will have twice bigger yield strength (we assume, as experimental data suggests, that the yield strength of graphene is much bigger than the transversal yield strength of Kevlar, and that linear elasticity approximation suffices for the estimate).

In Figure 7, we show the reduction of the maximum  value of the von Mises stress in the Kevlar fiber region, as a function of the ratio of  the Kevlar fiber radius $R_1 $ to the total radius $R_2,$ for various loading cases.

Figure 7 (a) shows a uni-axial compression in the $Y$ direction, applied to  the outer graphene layer,  and having a sharp peak of
%polar angle
angular width $\frac{\pi}{16} $ on the $Y$ axes. \mbox{Figure 7(a)} illustrates typical displacements for such load. In Figure 7(d), we show the reduction of the maximum  von Mises stress in the Kevlar fiber region, $\frac{\sigma_{\mbox{{\tiny max}}} \lbr R_2 \rbr }{\sigma_{\mbox{{\tiny max}}} \lbr R_1 \rbr}$  as a function of the inner to outer radius ratio $\frac{R_1}{R_2}.$

Figure 7 (b) shows an off-center compression in the $Y$ direction, applied to the outer graphene layer at 4 points, having  polar   angles $\pm \frac{\pi}{4},$  $\pm \frac{3 \pi}{4}.$  Each load has angular width $\frac{\pi}{16} .$
Figure 7 (c) shows a one-sided version of such off-center compression, applied at 2 points having polar   angles $\pm \frac{\pi}{4}.$ In Figures 7 (b), (c) we illustrate typical displacements for such loads. Reduction of the maximum  von Mises stress for such loads is shown in Figures 7 (e) and 7 (f),  respectively.

The calculation reveals that the reduction of the maximum  von Mises stress for all the three different loading cases was similar.

%(a), we plot   $\sigma_{\mbox{\small max }} (R_1) $ , the maximum  value of the Von Mises stress in the Kevlar fiber region as a function of the  Kevlar fiber radius $R_1 .$  The outer radius $R_2 $ of the Kevlar-graphene system stays fixed (in fact in continuous model stresses, including the maximum  value of the Von Mises stress, depend only  on the ratio $\frac{R_1}{R_2}$). The external load is uni-axial compression in the $Y$ direction, applied to  the outer graphene layer,  and having a sharp peak of polar angle width $\frac{\pi}{16} $ on the $Y$ axes; this load does not change throughout the  computation.

\section*{Conclusions}

Strengthening of Kevlar-graphene fibers, with the increase of thickness of the outer layer of graphene, was  predicted by analyzing  an explicit solution of the equations of 2D linear elasticity.

Results of computation of von Mises stress in Kevlar-graphene fibers for different transversal loadings demonstrated a two-fold increase in the yield strength, when the outer graphene shell radius was about 4 \% (corresponding to about 8 \% weight) of the  Kevlar fiber radius.

%We show that
%taking realistic Young modulus and Poisson ratios for transversal deformations, reinforcement starts to occur

The theoretical results presented here can potentially be used to guide experimental work, and to model strengthening of other fiber reinforced composites.

\section*{Acknowledgments}

We acknowledge  the financial support of this research by the National Science Foundation, grants 0933140 and 1138205.

\newpage
\section*{Figure Captions}

%\noindent\hspace{-14mm} \includegraphics[width= 7.5 in, height= 5.in]{fig1_.pdf}
%\\
{ {\bf Figure 1.} (a) - Schematic diagram of Kevlar-graphene system: the inner part is Kevlar fiber with radius $R_1,$ the outer layer is graphene with radius $R_2$: ($R_2>R_1$); (b) - Displacements on the whole domain, in the case $\frac{R_1}{R_2} =0.70$  }
\vspace{5mm}

%\newpage

%\noindent\hspace{-14mm}  \includegraphics[width= 8.625 in, height= 5.75 in]{fig2.pdf}
%\\

%\small
\noindent {\bf Figure 2.} (a)  Applied force per unit transverse area,  as a function of polar angle. Force is in the $Y$ direction, and is nonzero in an interval of width $\theta_w = \frac{\pi}{16}$ centered at $\theta=  \pm \dst\frac{\pi}{2};$ (b) $\sigma_{rr}$ component of material stress on the outer boundary, as a function of polar angle $\theta ;$
(c) $\sigma_{r\theta}$ component of material stress on the outer boundary, as a function of polar angle $\theta$ (polar coodinates are denoted $r, \theta$ )
\vspace{3mm}

%\newpage

%\noindent\hspace{-14mm}\includegraphics[width= 3.4in, height= 3 in]{fig3_a.pdf}\includegraphics[width= 3.4in, height= 3in]{fig3_b.pdf}

%\noindent\hspace{-14mm} \includegraphics[width= 3.4in, height= 3in]{fig3_c.pdf}\includegraphics[width= 3.4in, height= 3in]{fig3_d.pdf}

\noindent {
%\small
{\bf Figure 3.}   $\sigma_{yy}$ component of the stress on the Y axes (that is on the line where the load is applied), for varying values of the  ratio of the inner to outer radius  $\frac{R_1}{R_2}:$ (a)  $\frac{R_1}{R_2} =0.90$ (b) $\frac{R_1}{R_2} =0.93$ (c)  $\frac{R_1}{R_2} =0.96$  (d) $\frac{R_1}{R_2} =0.99$
%The outer radius stays at  $R_2= 100.$
Applied force per unit area is the same ,at $\pm 1$   unit, and is in the $Y$ direction; it is applied on the outer boundary at  polar angle  intervals of width $\frac{\pi}{16}$ centered at $ \theta = \pm \frac{\pi}{2}.$
\\
There is a marked drop of stress occurring in the outer reinforcing graphene  layer, provided that  $\frac{R_2 -R_1}{R_2}  \gtrsim 4 \% .$
}
\vspace{5mm}

%\newpage
%\mbox{}
%\vspace{-20mm}

%\noindent\hspace{-14mm}\includegraphics[width= 15.3 cm, height= 21 cm]{fig4.pdf}

%\vspace{-45mm}
\noindent
%\hspace{-14mm}
{
%\small
 {\bf Figure 4.}  $\sigma_{xx}$,$\sigma_{yy}$,$\sigma_{xy}$ components of the stress tensor  on the Kevlar fiber  domain, for  varying values of the  ratio of the inner to outer radius  $\frac{R_1}{R_2}:$
\\
(a) $\sigma_{xx},$   $\frac{R_1}{R_2} =0.90$  (b) $\sigma_{yy},$   $\frac{R_1}{R_2} =0.90$  (c) $\sigma_{xy} ,$   $\frac{R_1}{R_2} =0.90$
\\
(d) $\sigma_{xx},$   $\frac{R_1}{R_2} =0.93$ (e) $\sigma_{yy},$   $\frac{R_1}{R_2} =0.93$  (f) $\sigma_{xy} ,$   $\frac{R_1}{R_2} =0.93$
\\
(g) $\sigma_{xx},$   $\frac{R_1}{R_2} =0.96$  (h) $\sigma_{yy},$  $\frac{R_1}{R_2} =0.96$   (i) $\sigma_{xy} ,$  $\frac{R_1}{R_2} =0.96$
\\
(j) $\sigma_{xx},$  $\frac{R_1}{R_2} =0.99$   (k) $\sigma_{yy},$  $\frac{R_1}{R_2} =0.99$  (l) $\sigma_{xy} ,$  $\frac{R_1}{R_2} =0.99$ }
\\
%The outer radius stays at  $R_2= 100.$
Applied on the outer boundary  force per unit area is the same, at $\pm 1$   unit, and is in the $Y$ direction; it is applied on the outer boundary within a polar angle  interval  of width $\frac{\pi}{16}$ centered at $ \theta = \pm \frac{\pi}{2}.$
\vspace{5mm}
%\newpage
%\mbox{}
%\vspace{-20mm}

%\noindent\hspace{-14mm}\includegraphics[width= 15.3 cm, height= 21 cm]{fig5.pdf}

%\vspace{-45mm}
%\noindent\hspace{-14mm}
{
%\small
\noindent  {\bf Figure 5.} $\sigma_{rr}$ , $\sigma_{r\theta}$  polar coordinates components of the stress tensor  on the Kevlar fiber  domain, for  varying values of the  ratio of the inner to outer radius  $\frac{R_1}{R_2}:$
\\
(a) $\sigma_{rr}$ , $\frac{R_1}{R_2} =0.90$, (b) $\sigma_{r\theta}$ , $\frac{R_1}{R_2} =0.90$,
\\
(c) $\sigma_{rr}$ , $\frac{R_1}{R_2} =0.93$, (d) $\sigma_{r\theta}$ , $\frac{R_1}{R_2} =0.93$,
\\
(e) $\sigma_{rr}$ , $\frac{R_1}{R_2} =0.96$, (f) $\sigma_{r\theta}$ , $\frac{R_1}{R_2} =0.96$,
\\
(g) $\sigma_{rr}$ , $\frac{R_1}{R_2} =0.99$, (h) $\sigma_{r\theta}$ , $\frac{R_1}{R_2} =0.99$ .
\\
Applied on the outer boundary force per unit area is the same, at $\pm 1$   unit, and is in the $Y$ direction;  it is applied on the outer boundary within a polar angle  interval  of width $\frac{\pi}{16}$ centered at $ \theta = \pm \frac{\pi}{2}.$ }
\vspace{5mm}

%\newpage

%We now look at Von Mises  stress on the whole domain%, and also in the inner region of the fiber, with harder outer shell

%\includegraphics[width= 3.in, height= 2.4in]{fig6_a.pdf}
%\includegraphics[width= 3.in, height= 2.4in]{fig6_b.pdf}
%\\
{
%\small
\noindent {\bf Figure 6.} Von Mises  stress $\sigma_{M  }$  (a) on the whole domain , (b) in  the fiber region only, in the case where ratio of the inner to outer radius  $\frac{R_1}{R_2}$ is 0.95 .  Applied force per unit area is  $\pm 1$   unit, and is a compression along the $Y$ axes; it is applied on the outer boundary within a polar angle  interval  of width $\frac{\pi}{16}$ centered at $ \theta =  \frac{\pi}{2},$ with an opposite force applied in the diametrically opposite region.  }
\vspace{5mm}

%\newpage

%\noindent\hspace{-18mm}\includegraphics[width= 7.8in, height= 6.2in]{fig7.pdf}
%\\
{
%\small
\noindent {\bf Figure 7.}  External loads, corresponding typical displacements, and reductions of the maximum  Von Mises  stress in the Kevlar fiber region  as   a function of the ratio of the  Kevlar fiber radius $R_1 $ to the total radius $R_2:$
\\
(a) uni-axial compression in the $Y$ direction, applied to  the outer graphene layer,  and having a sharp peak of angular width $\frac{\pi}{16} $ on the $Y$ axes. Typical displacements  are shown.
\\
(b)off-center compression in the $Y$ direction, applied to the outer graphene layer at 4 points, having  polar   angles $\pm \frac{\pi}{4},$  $\pm \frac{3 \pi}{4},$ and   of angular width  $\frac{\pi}{16}$ each .%Typical displacements  are shown.
\\
(c) one-sided off-center compression in the $Y$ direction, applied to the outer graphene layer at two  points, having polar   angles $\pm \frac{\pi}{4},$ and   of angular width  $\frac{\pi}{16}$ each. %Typical displacements  are shown.
\\
(d) reduction of the maximum  Von Mises  stress in the Kevlar fiber region, as   a function of the ratio of the  Kevlar fiber radius $R_1 $ to the total radius $R_2,$ for the central loading shown in Fig. 7(a).
\\
(e) reduction of the maximum  Von Mises  stress in the Kevlar fiber region, as   a function of $\frac{R_1}{R_2},$ in the case  of symmetric off-center loading,  Fig. 7(b)
\\
(f)reduction of the maximum  Von Mises  stress in the Kevlar fiber region, as   a function of $\frac{R_1}{R_2},$ in the case  of one-sided off-center loading, Fig. 7(c).
\\
Reference red line in (d), (e), (f) corresponds to no graphene reinforcement.

\newpage

\noindent\hspace{-14mm} \includegraphics[width= 7.5 in, height= 5.in]{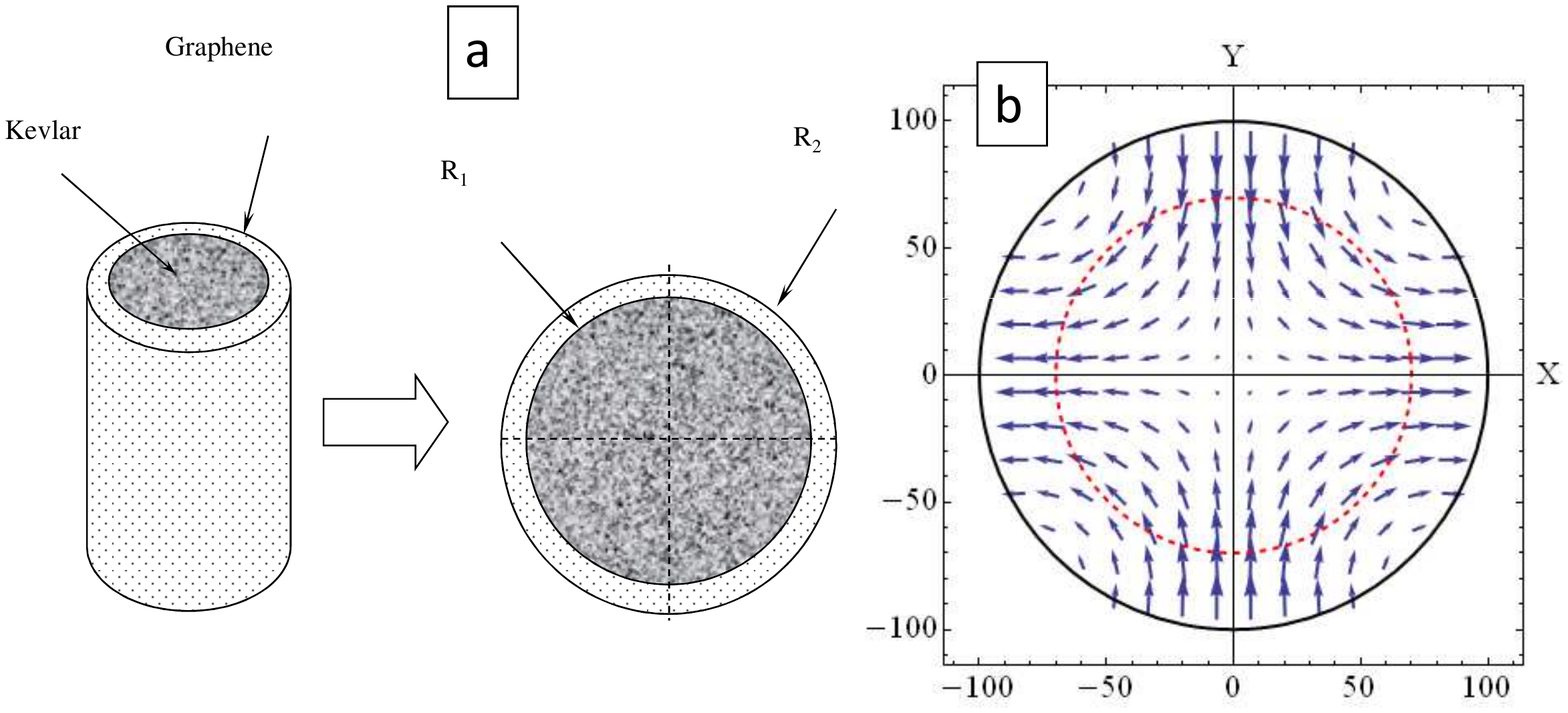}
\\
{\small {\bf Figure 1.} (a) - Schematic diagram of Kevlar-graphene system: the inner part is Kevlar fiber with radius $R_1,$ the outer layer is graphene with radius $R_2$: ($R_2>R_1$); (b) - Displacements on the whole domain, in the case $\frac{R_1}{R_2} =0.70$  }
\vspace{3mm}

\newpage

\noindent\hspace{-14mm}  \includegraphics[width= 8.625 in, height= 5.75 in]{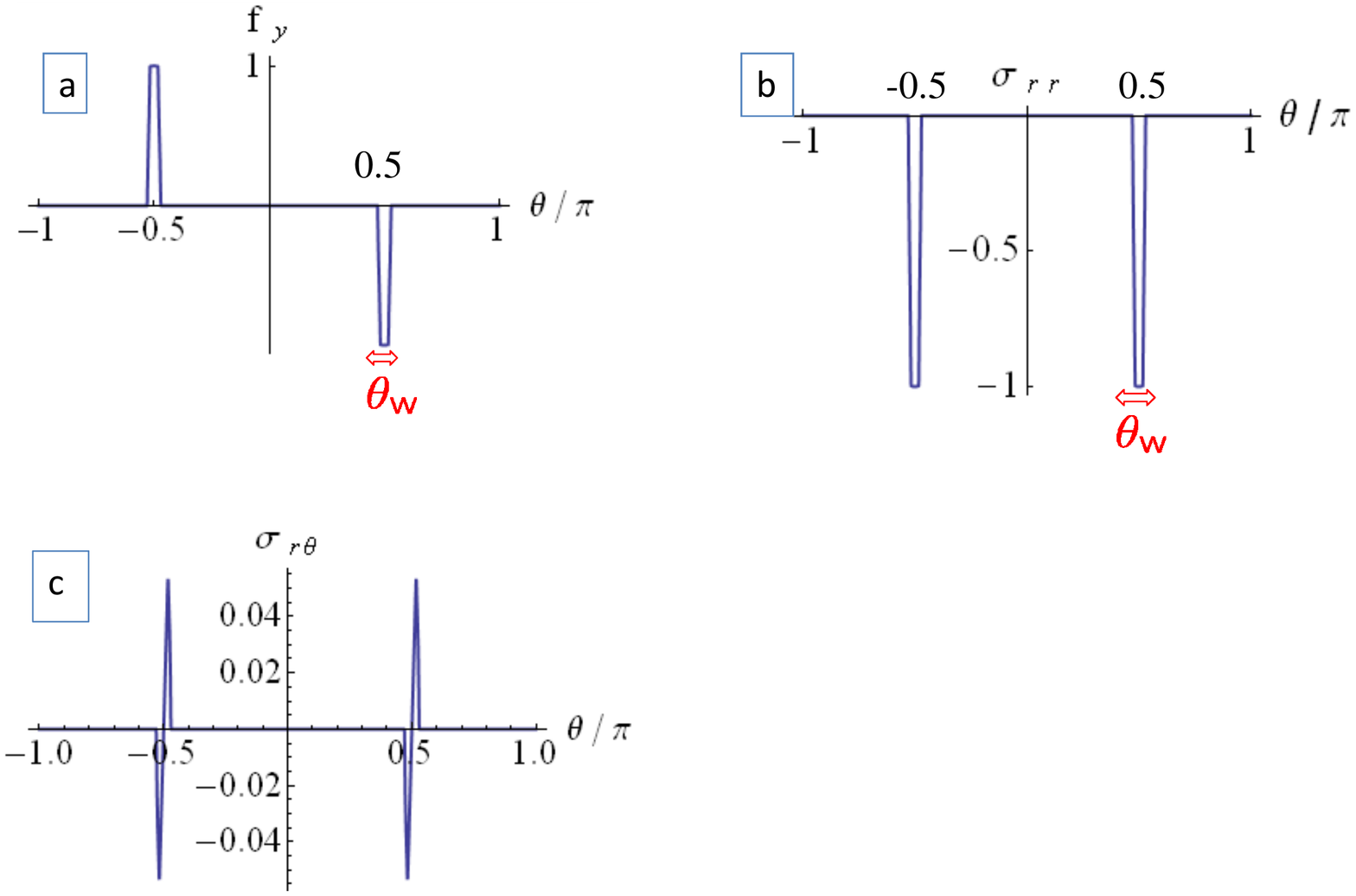}
\\
{\small {\bf Figure 2.} (a)  Applied force per unit transverse area,  as a function of polar angle. Force is in the $Y$ direction, and is nonzero in an interval of width $\theta_w = \frac{\pi}{16}$ centered at $\theta=  \pm \dst\frac{\pi}{2};$ (b) $\sigma_{rr}$ component of material stress on the outer boundary, as a function of polar angle $\theta ;$
(c) $\sigma_{r\theta}$ component of material stress on the outer boundary, as a function of polar angle $\theta$ (polar coodinates are denoted $r, \theta$ )
\vspace{3mm}

\newpage

\noindent\hspace{-14mm}\includegraphics[width= 3.4in, height= 3 in]{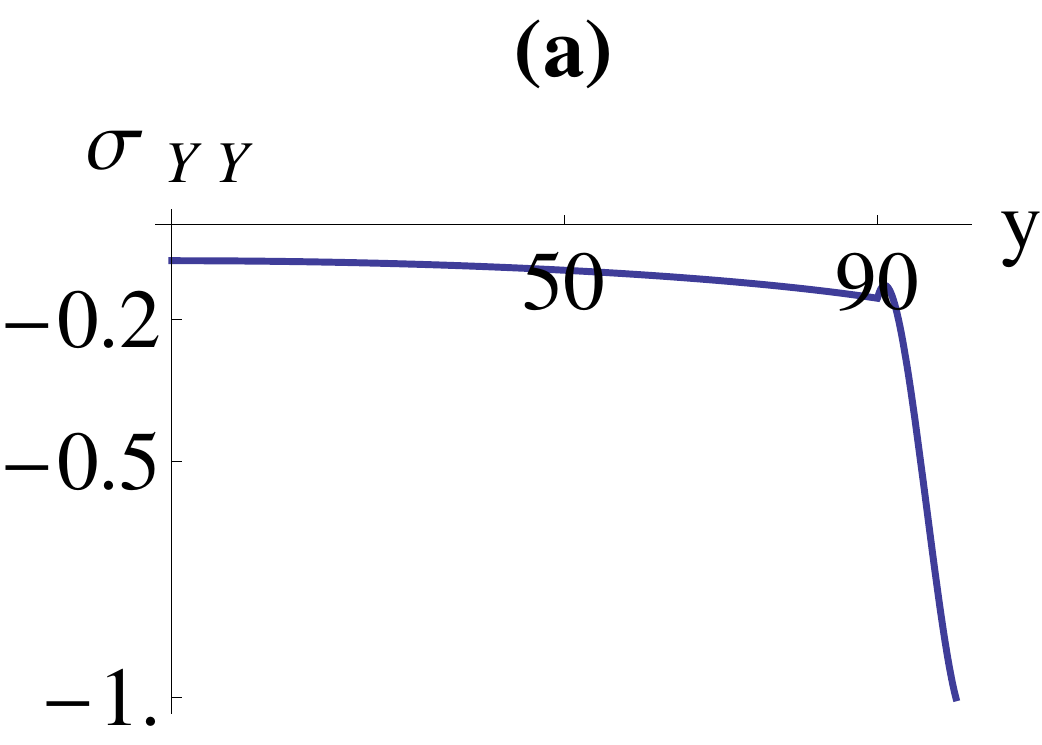}\includegraphics[width= 3.4in, height= 3in]{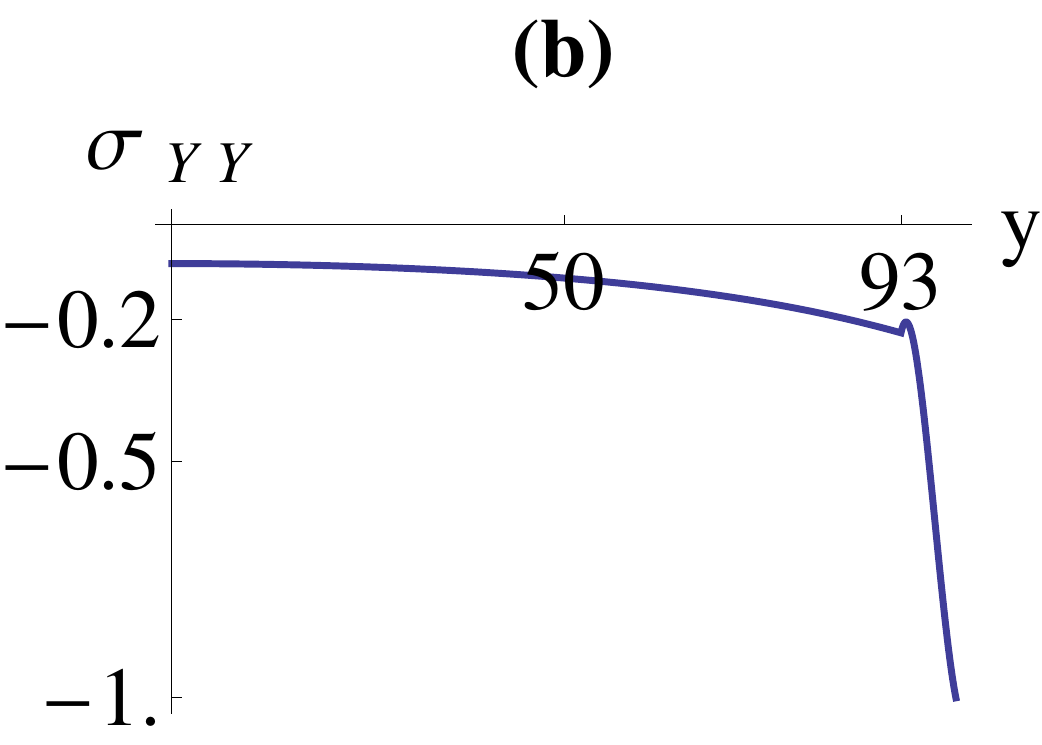}

\noindent\hspace{-14mm} \includegraphics[width= 3.4in, height= 3in]{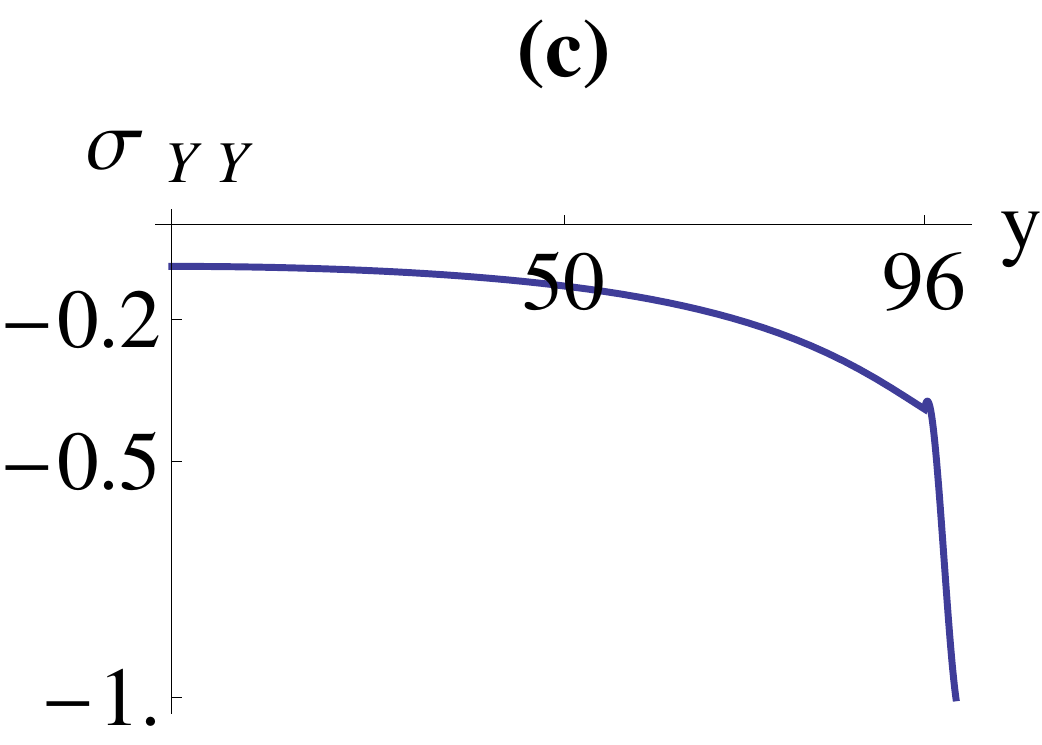}\includegraphics[width= 3.4in, height= 3in]{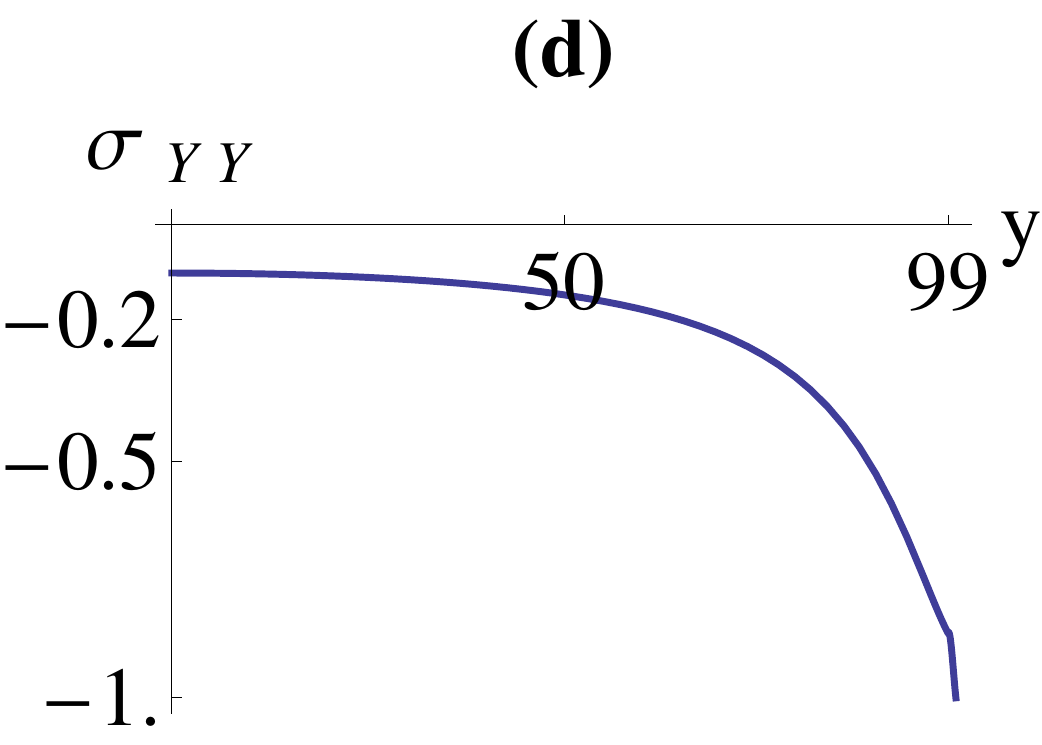}

{\small {\bf Figure 3.}   $\sigma_{yy}$ component of the stress on the Y axes (that is on the line where the load is applied), for varying values of the  ratio of the inner to outer radius  $\frac{R_1}{R_2}:$ (a)  $\frac{R_1}{R_2} =0.90$ (b) $\frac{R_1}{R_2} =0.93$ (c)  $\frac{R_1}{R_2} =0.96$  (d) $\frac{R_1}{R_2} =0.99$
%The outer radius stays at  $R_2= 100.$
Applied force per unit area is the same ,at $\pm 1$   unit, and is in the $Y$ direction; it is applied on the outer boundary at  polar angle  intervals of width $\frac{\pi}{16}$ centered at $ \theta = \pm \frac{\pi}{2}.$
\\
There is a marked drop of stress occurring in the outer reinforcing graphene  layer, provided that  $\frac{R_2 -R_1}{R_2}  \gtrsim 4 \% .$
}

\newpage
\mbox{}
\vspace{-20mm}

\noindent\hspace{-14mm}\includegraphics[width= 15.3 cm, height= 21 cm]{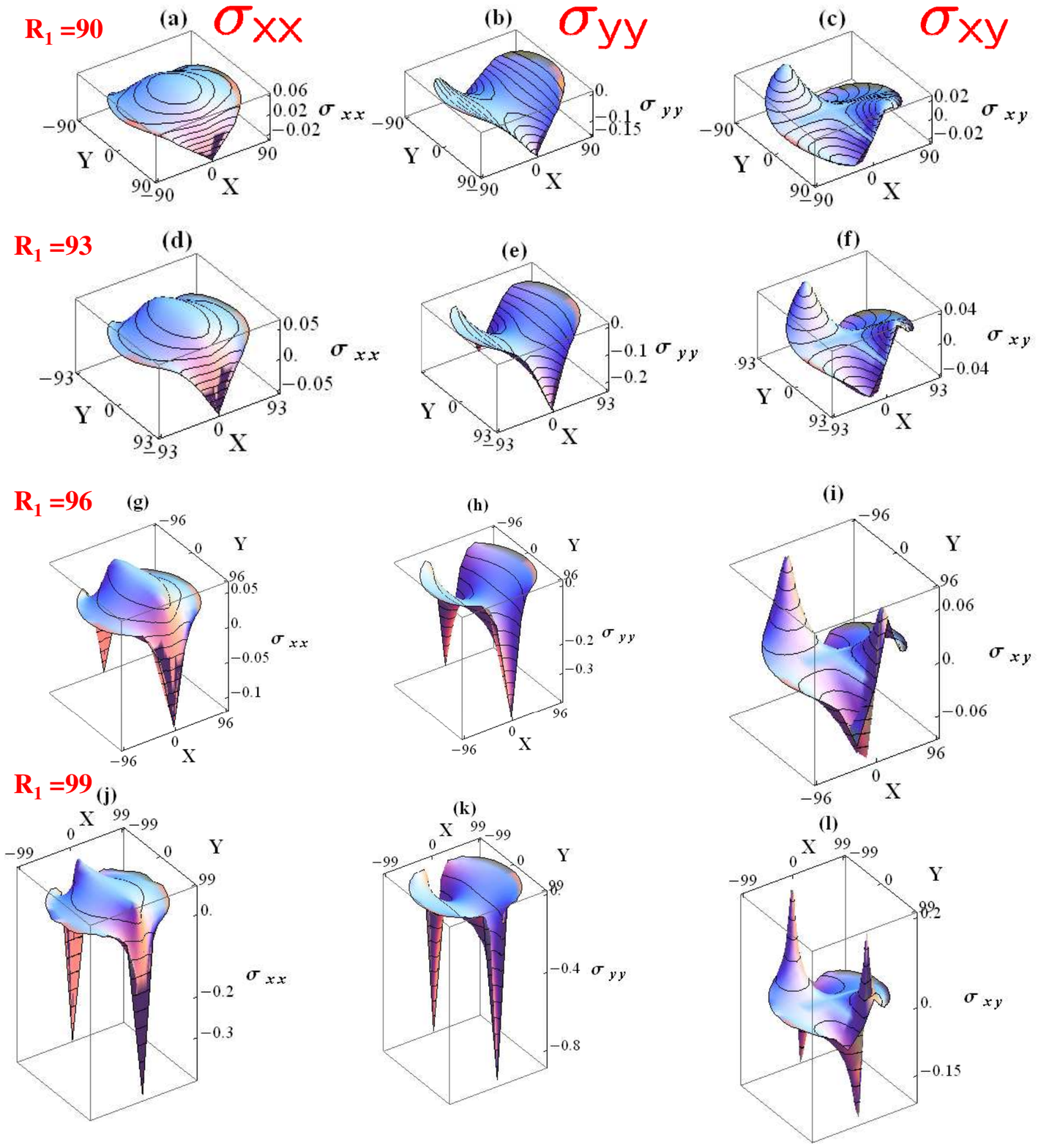}

\vspace{-45mm}
\noindent\hspace{-14mm}{\small  {\bf Figure 4.}  $\sigma_{xx}$,$\sigma_{yy}$,$\sigma_{xy}$ components of the stress tensor  on the Kevlar fiber  domain, for  varying values of the  ratio of the inner to outer radius  $\frac{R_1}{R_2}:$
\\
(a) $\sigma_{xx},$   $\frac{R_1}{R_2} =0.90$  (b) $\sigma_{yy},$   $\frac{R_1}{R_2} =0.90$  (c) $\sigma_{xy} ,$   $\frac{R_1}{R_2} =0.90$
\\
(d) $\sigma_{xx},$   $\frac{R_1}{R_2} =0.93$ (e) $\sigma_{yy},$   $\frac{R_1}{R_2} =0.93$  (f) $\sigma_{xy} ,$   $\frac{R_1}{R_2} =0.93$
\\
(g) $\sigma_{xx},$   $\frac{R_1}{R_2} =0.96$  (h) $\sigma_{yy},$  $\frac{R_1}{R_2} =0.96$   (i) $\sigma_{xy} ,$  $\frac{R_1}{R_2} =0.96$
\\
(j) $\sigma_{xx},$  $\frac{R_1}{R_2} =0.99$   (k) $\sigma_{yy},$  $\frac{R_1}{R_2} =0.99$  (l) $\sigma_{xy} ,$  $\frac{R_1}{R_2} =0.99$ }
\\
%The outer radius stays at  $R_2= 100.$
Applied on the outer boundary  force per unit area is the same, at $\pm 1$   unit, and is in the $Y$ direction; it is applied on the outer boundary within a polar angle  interval  of width $\frac{\pi}{16}$ centered at $ \theta = \pm \frac{\pi}{2}.$
\newpage
\mbox{}
\vspace{-20mm}

\noindent\hspace{-14mm}\includegraphics[width= 15.3 cm, height= 21 cm]{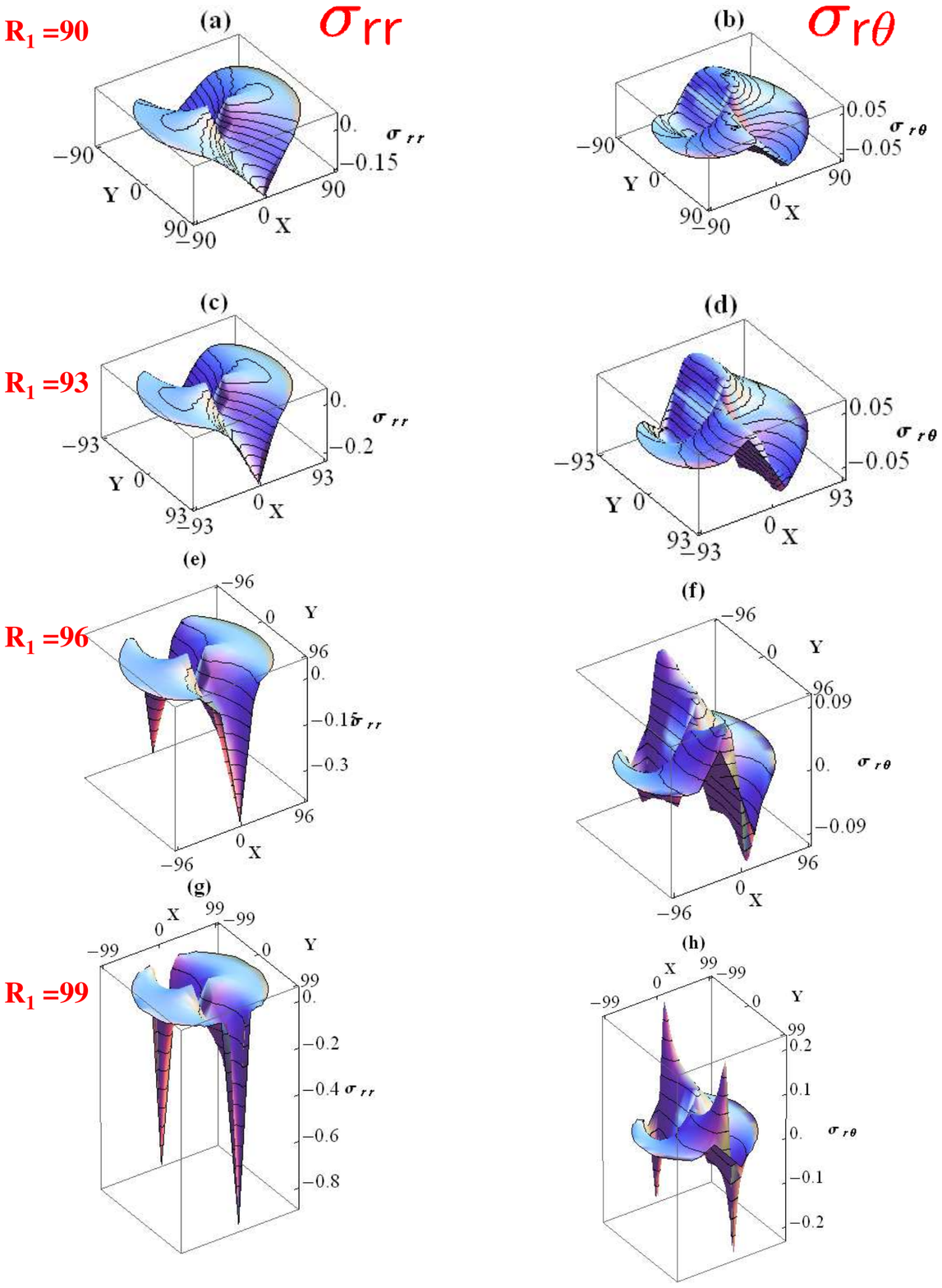}

\vspace{-45mm}
\noindent\hspace{-14mm}
{\small {\bf Figure 5.} $\sigma_{rr}$ , $\sigma_{r\theta}$  polar coordinates components of the stress tensor  on the Kevlar fiber  domain, for  varying values of the  ratio of the inner to outer radius  $\frac{R_1}{R_2}:$
\\
(a) $\sigma_{rr}$ , $\frac{R_1}{R_2} =0.90$, (b) $\sigma_{r\theta}$ , $\frac{R_1}{R_2} =0.90$,
\\
(c) $\sigma_{rr}$ , $\frac{R_1}{R_2} =0.93$, (d) $\sigma_{r\theta}$ , $\frac{R_1}{R_2} =0.93$,
\\
(e) $\sigma_{rr}$ , $\frac{R_1}{R_2} =0.96$, (f) $\sigma_{r\theta}$ , $\frac{R_1}{R_2} =0.96$,
\\
(g) $\sigma_{rr}$ , $\frac{R_1}{R_2} =0.99$, (h) $\sigma_{r\theta}$ , $\frac{R_1}{R_2} =0.99$ .
\\
Applied on the outer boundary force per unit area is the same, at $\pm 1$   unit, and is in the $Y$ direction;  it is applied on the outer boundary within a polar angle  interval  of width $\frac{\pi}{16}$ centered at $ \theta = \pm \frac{\pi}{2}.$ }

\newpage

%We now look at Von Mises  stress on the whole domain%, and also in the inner region of the fiber, with harder outer shell

\includegraphics[width= 3.in, height= 2.4in]{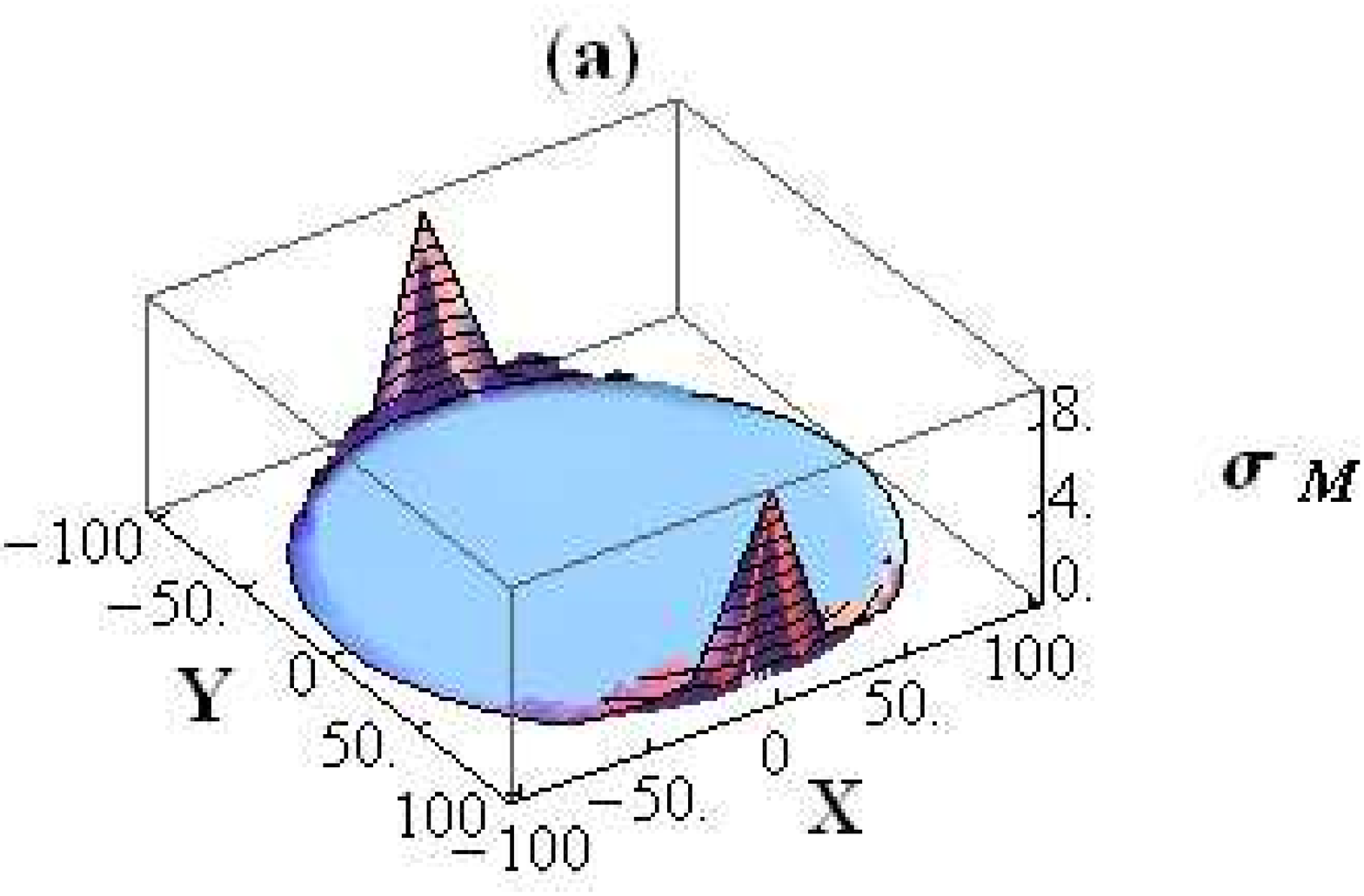}
\includegraphics[width= 3.in, height= 2.4in]{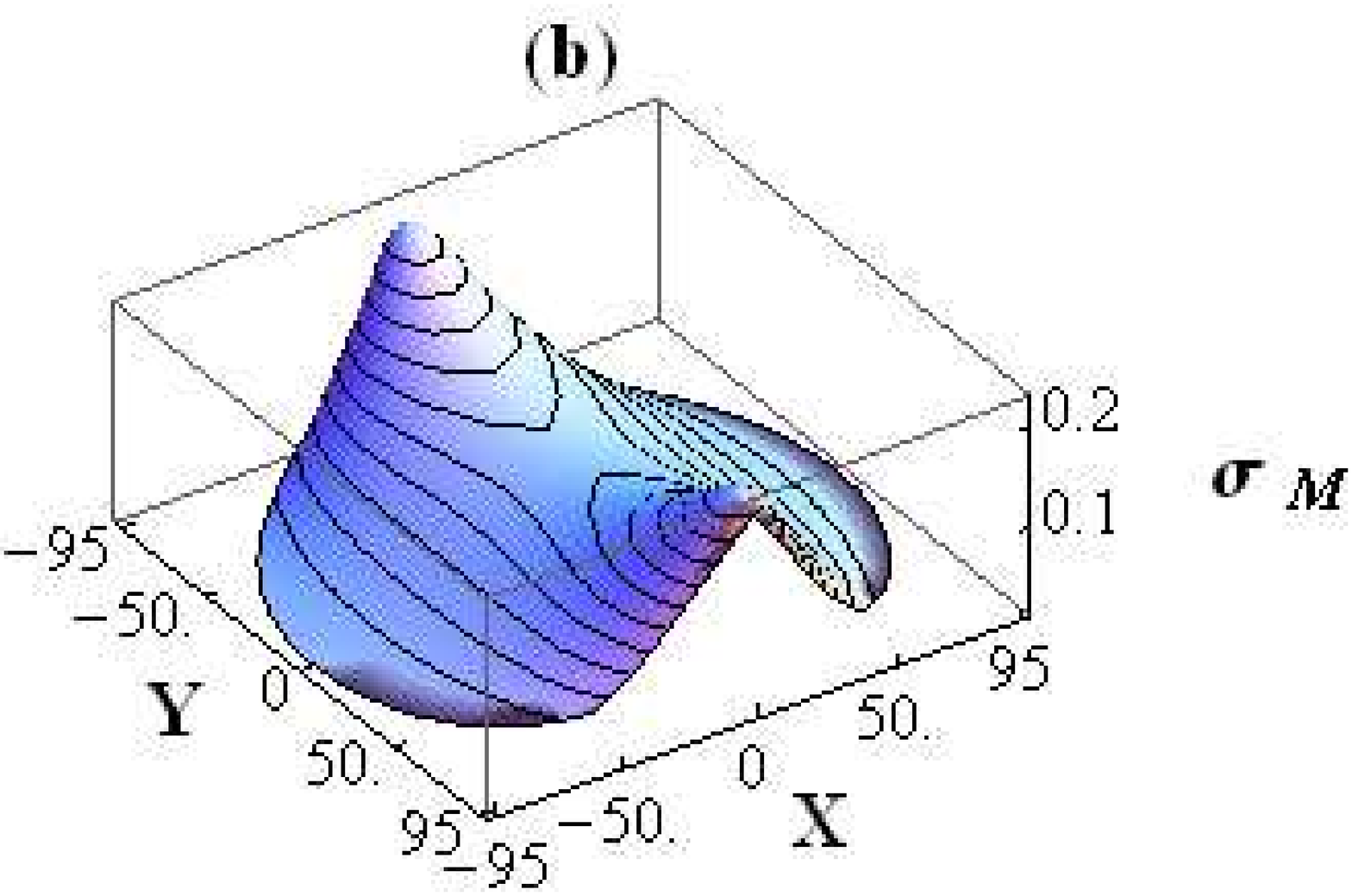}
\\
{\small {\bf Figure 6.} Von Mises  stress $\sigma_{M  }$  (a) on the whole domain , (b) in  the fiber region only, in the case where ratio of the inner to outer radius  $\frac{R_1}{R_2}$ is 0.95 .  Applied force per unit area is  $\pm 1$   unit, and is a compression along the $Y$ axes; it is applied on the outer boundary within a polar angle  interval  of width $\frac{\pi}{16}$ centered at $ \theta =  \frac{\pi}{2},$ with an opposite force applied in the diametrically opposite region.  }

\newpage

\noindent\hspace{-18mm}\includegraphics[width= 8.4in, height= 6.3in]{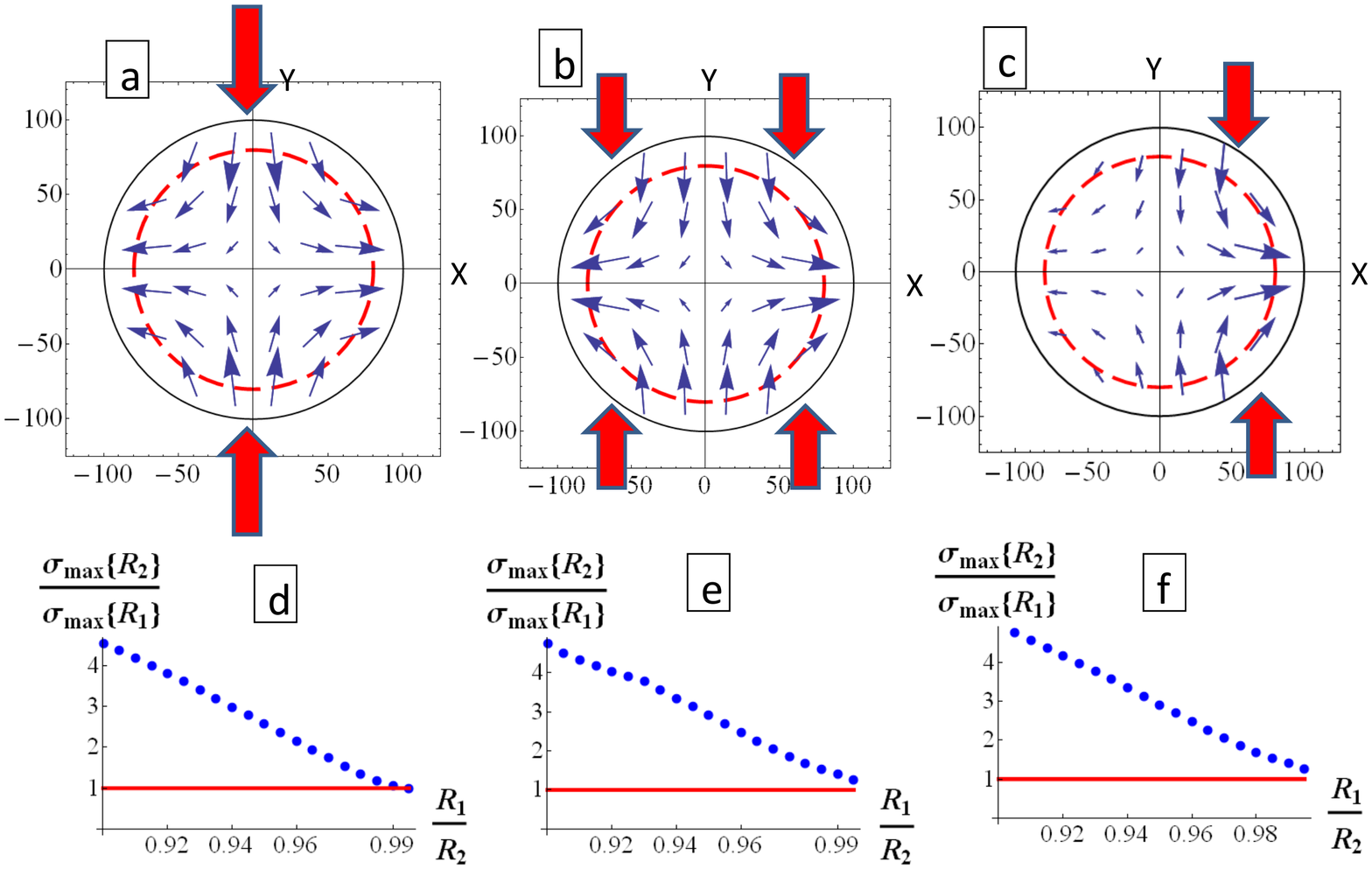}
\\
{\small
\noindent {\bf Figure 7.}   External loads, corresponding typical displacements, and reductions of the maximum  Von Mises  stress in the Kevlar fiber region  as   a function of the ratio of the  Kevlar fiber radius $R_1 $ to the total radius $R_2:$
\\
(a) uni-axial compression in the $Y$ direction, applied to  the outer graphene layer,  and having a sharp peak of angular width $\frac{\pi}{16} $ on the $Y$ axes. Typical displacements  are shown.
\\
(b)off-center compression in the $Y$ direction, applied to the outer graphene layer at 4 points, having  polar   angles $\pm \frac{\pi}{4},$  $\pm \frac{3 \pi}{4},$ and   of angular width  $\frac{\pi}{16}$ each .%Typical displacements  are shown.
\\
(c) one-sided off-center compression in the $Y$ direction, applied to the outer graphene layer at two  points, having polar   angles $\pm \frac{\pi}{4},$ and   of angular width  $\frac{\pi}{16}$ each. %Typical displacements  are shown.
\\
(d) reduction of the maximum  Von Mises  stress in the Kevlar fiber region, as   a function of the ratio of the  Kevlar fiber radius $R_1 $ to the total radius $R_2,$ for the central loading shown in Fig. 7(a).
\\
(e) reduction of the maximum  Von Mises  stress in the Kevlar fiber region, as   a function of $\frac{R_1}{R_2},$ in the case  of symmetric off-center loading,  Fig. 7(b)
\\
(f)reduction of the maximum  Von Mises  stress in the Kevlar fiber region, as   a function of $\frac{R_1}{R_2},$ in the case  of one-sided off-center loading, Fig. 7(c).
\\
Reference red line in (d), (e), (f) corresponds to no graphene reinforcement.

%(a) Blue dots: the maximum  value of the von Mises stress in the  Kevlar fiber region, when the inner Kevlar region has radius $R_1$, and the graphene shell has a fixed outer  radius $R_2=100.$    Reference red line correspond to no graphene reinforcement. Applied on the outer boundary force per unit area is always the same, at $\pm 1$   unit, and is in the $Y$ direction; it is applied on the outer boundary within a polar angle  interval  of width $\frac{\pi}{16}$ centered at $ \theta = \pm \frac{\pi}{2}.$
%\\
%(b) $\dst\frac{\sigma_{\mbox{\small max }} (R_2)} {\sigma_{\mbox{\small max }} (R_1)}   ,$  a factor of the  increase in the yield strength of Kevlar-graphene system,
%as a function of
% $\dst\frac{R_1}{R_2}.$

}
\vspace{3mm}
%\end{figure}

\end{document}